\documentstyle[]{article}
\begin{document}
\title{Super black hole as spinning particle: Supersymmetric baglike core
\footnote{Plenary talks given at the School-Workshop Praha-Spin-2001
(Prague,July 15-28,2001) and at the XXIV International Workshop on
Fundamental Problems of HEP and Field Theory (IHEP, June 2001, Protvino).}}
\author{Alexander Burinskii \thanks{e-mail: bur@ibrae.ac.ru} \\
Gravity Research Group, NSI, Russian Academy of Sciences}

\maketitle

\begin{abstract}
We consider  particlelike solutions to
supergravity based on the Kerr-Newman black hole (BH) solution.
The BH singularity is regularized by means of a phase transition to a new
vacuum state near the core region confining a dual gauge field.

Supersymmetric BPS-saturated domain wall model is suggested which can
provide this phase transition and formation the stable charged
superconducting core.

  For spinning particle the core takes the form of thin,
relativistically rotaiting disk.

\end{abstract}

\section{Introduction}     

Properties of non-perturbative quantum models are essentially determined
by the properties of underlying classical models. The Kerr-Newman BH
solution seems the most suitable classical background for spinning
particles.
Supersymmetry and supergravity give extra advantages leading to cancelation
of quantum divergences. Moreover, the solutions saturating BPS-bound  and
retaining a part of supersymmetry may not receive quantum loop corrections.

\par
In 1969 Carter observed \cite{Car},  that  if three parameters of the
Kerr-Newman solution are adopted to be ($\hbar $=c=1 )
\begin{equation}
\quad e^{2}\approx  1/137,\quad
m \approx 10^{-22},\quad a \approx  10^{22},\quad ma=1/2,
\label{par}
\end{equation}
then one obtains  a  model  for  the four  parameters  of  the electron:
charge $e$, mass $m$, spin $l$ and  magnetic moment $ea$,
and the gyromagnetic ratio is automatically the same as that of the Dirac
electron.  The first treatment of the source of the Kerr spinning
particle was given by Israel \cite{Isr} in the form of an infinitely thin
disk spanned by the Kerr singular ring. Disk has the Compton size with radius
$a=l/m=\frac 1{2m}$. The Israel results where corrected by Hamity
showing that the disk is to be in a rigid relativistic rotation, and
L\`opez suggested a regularized model of the source
in the form of a  rotating ellipsoidal shell ( bubble ) covering the
singular ring \cite{Lop}.
The structure of the electromagnetic field near the disk suggested
superconducting properties of the material of the source,
and there was obtained an analogue of the Kerr singular ring with
the Nielsen-Olesen and Witten superconducting strings.
Since 1992 there has been considerable interest as to black holes in
superstring theory, and the point of view appeared that some of black holes
can be treated as elementary particles \cite{part}.
  In particular, Sen \cite{Sen} has obtained a generalization of the Kerr
solution to low energy string theory, and it was shown \cite{BS} that near
the Kerr singular ring the Kerr-Sen solution acquires a metric similar to
the field around a heterotic string.

The simplest consistent Super-Kerr-Newman
BH solution \cite{SBH} was constructed on the base of the
(broken) Ferrara-Nieuvenhuisen N=2 Einstein-Maxwell D=4 supergravity.
Since source (or singularity) of this solution is covered by BH horizon, the
matter chiral fields of supergravity are not involved at all.
However, for the large angular momentum corresponding to spinning particles
the Kerr horizons are absent, and there appears a naked singularity.
It can be regularized by a matter source \cite{Bag} built of the nontrivial
chiral (Higgs) fields.

   One of the approaches to regularization of the particlelike BH solutions
is based on the old idea of the replacement of singularity by a "semiclosed
world", internal space-time of a constant curvature (M. Markov, 1965;
I. Dymnikova \cite{Dym}).
\footnote{The Dirac classical electron model (generalization of this model
to the charged and rotating bubble in gravity was given by L\`opez, 1984,
see ref. in \cite{Bag}), as well as the bag models could also be included
in this class when one assumes that regularization is provided by a flat
core region.}

We consider development of these models leading to a  non-perturbative
soliton-like solution to supergravity and assuming that the external field
is the Kerr-Newman black hole solution, and the core is described by a domain
wall bubble based on the chiral fields of a supersymmetric field model.

\section{Regular sources for the rotating and non-rotating black hole
solutions of the Kerr-Schild class}

The Kerr-Schild class of metrics
\begin{equation}
g_{\mu \nu} = \eta_{\mu \nu} + 2h K_{\mu}K_{\nu}.
\label{gKS}
\end{equation}
allows one to consider the above regularization
for the rotating and nonrotating, charged and uncharged BH's in unique manner
\cite{Bag}. It allows one to describe the external BH field and the internal
(A)dS region, as well as a smooth interpolating region between them without
especial matching conditions, by using one smooth
function, $f(r)$, of the Kerr radial coordinate $r$.

Here $\eta$ is an
auxiliary Minkowski metric, $K_\mu$ is a vortex field of the Kerr
principal null congruence, and scalar function $h$ has the form
\footnote{For $a\ne 0$ the Kerr coordinates $r$ and $\theta$ are oblate
spheroidal ones.}
\begin{equation}
h=f(r)/(r^2 +a^2\cos ^2 \theta).
\label{hf}
\end{equation}
 In particular, for the Kerr-Newman BH solution
\begin{equation}
f(r)=f_{ext}(r)=mr-e^2/2,
\label{fKN}
\end{equation}
where $m$ and $e$ are the total mass and charge.
 The transfer to nonrotating case occurs by $a=0$ when the Kerr
congruence turns into a twist-free "hedgehog" configuration, and $r, \theta$
are usual spherical coordinates.
It is important that function $f(r)$ is not affected by this transfer
that allows one to simplify treatment concentrating on the $a=0$ case.

By $a=0$, the regularizing core region of a constant curvature can be
described by $f=f_{int}(r)= \alpha r^4 $, where $\alpha=\Lambda/6 $,
$\Lambda$ is cosmological constant, and energy density in core is
\footnote{In this case, as shows (\ref{hf}),
gravitational singularity is regularized also by $a\ne 0$.}
$\rho= \frac 3 4 \alpha / \pi$ .

A smooth matching of the internal and external metrics is provided by
smooth function $f(r)$ interpolating between $f_{int}$ and $f_{KN}$.
The radial position $r_0$ of the phase transition region can be estimated
as a point of intersection of the plots $f_{int}$ and $f_{ext}$,
\begin{equation}
\frac 4 3 \pi \rho r_0 ^4=mr_0-e^2/2.
\label{r0}
\end{equation}

Analysis shows \cite{Bag} that for charged sources there appears a
thin intermediate shell at $r= r_0$ with a strong tangential stress that
is typical for a domain wall structure.
Dividing this equation on $r_0$ one can recognize here the {\it mass balance
equation}
\begin{equation}
m = M_{int}(r_0) + M_{em}(r_0),
\label{mb}
\end{equation}
where $m$ is total mass, $M_{int}(r_0) $ is ADM mass of core and
$ M_{em}(r_0) = e^2/2r_0 $ is ADM mass of the external e.m. field.
It should be mentioned, that gravitational field is extremely small at $r_0$,
especially as $r_0$ is much more of gravitational radius
\footnote{In particular, if interior is flat ($\rho =0$ ) $r_0= e^2/2m$
-`classical electromagnetic radius'.}
( $r_0/m \sim 10^{42}$). Nevertheless, eq.
(\ref{mb}) shows that {\it phase transition is controlled by
gravity, but nonlocally!}
Note, that $M_{int}$ can be either
positive (that corresponds to dS interior) or negative ( AdS interior ).
As we shall see, supergravity suggests AdS vacua inside the bubble.

As consequence of this treatment we obtain also some demands to the
supergravity matter field model.

i - It has to provide a phase transition between internal
and external vacua.

ii - External vacuum has to be (super)-Kerr-Newman black hole solution
with {\it long range } electromagnetic field and  zero cosmological constant.

iii -Internal vacuum has to be (A)dS space with superconducting properties.

These demands are very restrictive and cannot satisfied in the known
    solitonlike bag, domain wall and bubble models.  Main contradiction is
connected with demands ii) and iii) since in the most of models external
electromagnetic field is short range. An exclusion is the
$U(I) \times \tilde U(I) $ field model which was used by Witten to describe
the cosmic superconducting strings \cite{Wit}. Our suggestion is to use
this field model for description the superconducting baglike configuration.

\section{Supersymmetric superconducting bag model}

The model contains two Higgs sectors: A and B. The chiral field of sector A,
$\phi (r)$ forms a structure similar to "lumps", Q-balls and the other known
non-topological solitons \footnote{See for example \cite{Col}}.
However, it has a specific potential
determined by a supersymmetric domain wall model, and it interacts with the
gauge field $A_{\mu}$, which forms the external long range electromagnetic
field $F_{\mu \nu}$.

Chiral field of sector B, $\sigma (r)$, forms a superconducting bag
confining the gauge field $B_{\mu}$,
( $F_{B\mu\nu} = B_{\mu , \nu} - B_{\nu , \mu}$).
There are some hints in favour of the dual superconductivity
for sector B.

     Supersymmetric version of the Witten field model (suggested by
J. Morris \cite{Mor}) has effective Lagrangian of the form
\begin{eqnarray}
L=-2(D^\mu \phi )\overline {( D_\mu  \phi  )}-2(\tilde D^\mu \sigma )
(\overline {\tilde D_\mu \sigma} )-\partial ^\mu Z \partial _\mu
\bar Z \nonumber \\
-\frac 14F^{\mu \nu }F_{\mu \nu }-
\frac 14 F_B ^{\mu \nu }F_{B\mu \nu}-V (\sigma, \phi, Z),
\label{bur-SL}
\end{eqnarray}
where $D_{\mu} = \nabla _{\mu} + ie A_{\mu}$,
$\tilde D_{\mu} = \nabla _{\mu} + ig B_{\mu}.$
The potential $V$ is determined through
the superpotential $W$ as
\begin{equation}
V= \sum _{i=1} ^5 \vert\partial _i W \vert^2,
\label{SV}
\end{equation}
and the superpotential $W(\Phi^i)$ is a holomorphic function of
the fife complex chiral fields
$\Phi ^i = \lbrace Z,\phi, \bar\phi, \sigma, \bar \sigma \rbrace $,
\begin{equation}
W=\lambda {Z}(\sigma \bar\sigma -\eta ^2) + ( c Z+m ) \phi\bar \phi.
\label{bur-SW}
\end{equation}
\par
In the effective Lagrangian the "bar" is identified with complex
conjugation, so there are really only three independent scalar fields,
and the
"new" ( neutral ) fields Z provides the synchronization of the phase
transition.
The supersymmetric vacuum states corresponding to the lowest value of the
potential are determined by the conditions
\begin{eqnarray}
 \partial _i W  =0;
\label{Svac}
\end{eqnarray}
and yield $V=0$.
These equations lead to two supersymmetric vacuum states:
\begin{equation}
I ) \qquad Z=0;\quad \phi=0 ;\quad \vert\sigma\vert=\eta ;\quad W=0,
\label{bur-true}
\end{equation}
we set it for external vacuum; and
\begin{equation}
II ) \qquad Z=-m/c;\quad \sigma=0; \quad \vert \phi \vert =\eta
\sqrt{\lambda/c};\quad W=\lambda m\eta ^2/c,
\label{bur-false}
\end{equation}
we set it as a state inside the bag.
\par

The treatment of the gauge field $A_\mu$ and $B_\mu$ in $B$-sector is similar
in many respects because of the symmetry between  $A$ and $B$ sectors
allowing one to consider the state $\Sigma = \eta$ in outer region as
superconducting one \footnote{The version of dual superconductivity
in B-sector seems the most interesting.}
in respect to the gauge field $B_\mu$.
Field $B_\mu$ acquires the mass $m_B= g \eta $ in outer region, and the
$\tilde U(I)$ gauge symmetry is broken, which provides confinement of the
$B_{\mu}$ field inside the bag.
The bag can also be filled by quantum excitations of fermionic,
or non Abelian fields.
\par
One can check the phase transition in the planar wall approximation
 ( neglecting the gauge fields ). It can be shown that it is a
BPS-saturated
domain wall solution interpolating between supersymmetric vacua I) and II).
Using the Bogomol'nyi
transformation one can represent the energy density as follows
\begin{eqnarray}
\rho &=& T_{00} =\frac{1}{2} \delta _{ij}\lbrack ( \Phi ^i,_z)
(\Phi ^j,_z) + (\frac {\partial W}{\partial \Phi ^i})
(\frac {\partial W}{\partial \Phi ^j})\rbrack \\
 &=&\frac{1}{2} \delta _{ij}\lbrack  \Phi ^i,_z + \frac {\partial W}{\partial
\Phi ^j}\rbrack
\lbrack  \Phi ^j,_z + \frac {\partial W}{\partial\Phi ^i}\rbrack
- \frac{\partial W}{\partial \Phi ^i}\Phi ^i,_z,
\label{trbog}\end{eqnarray}
where the last term is full derivative.
Then, integrating over the wall depth $z$ one obtains for the surface
energy density of the wall
\begin{equation}
\epsilon=\int _0 ^\infty \rho dz =\frac 12 \int\Sigma _i
( \Phi ^i,_z + \frac {\partial W}{\partial\Phi ^i})^2 dz + W(0) -W(\infty).
\label{edens}
\end{equation}
The minimum of energy is achieved when the first-order Bogomol'nyi
equations $\Phi ^i,_z + \frac {\partial W}{\partial\Phi ^i} =0$ are
satisfied, or in terms of $Z, \Phi, \Sigma $
\begin{eqnarray}
 Z^\prime &=& -  \lambda(\Sigma ^2-\eta ^2) - c \Phi ^2, \\
 \Sigma ^\prime &=& -\lambda Z\Sigma, \\
\Phi ^\prime &=& - (cZ +m) \Phi.
\label{bogeq}
\end{eqnarray}
Its value is given by
$\epsilon =W(0)-W(\infty)=\lambda m \eta^2/c$.
Therefore, this domain wall is BPS-saturated solution.
One can see that
the field $Z$, which appears only in the supersymmetric version
of the model, plays an essential role for formation of the phase transition.
\par
The structure of stress-energy tensor contains the typical for domain walls
tangential stress. The non-zero components of the stress-energy tensor have
the form
\begin{eqnarray}
T_{00} & = &-T_{xx} =- T_{yy}=\frac{1}{2}[ \delta _{ij}( \Phi ^i,_z)
(\Phi ^j,_z) + V];\\
T_{zz}& = &\frac{1}{2}[ \delta_{ij}( \Phi ^i,_z)(\Phi ^j,_z) - V] =0.
\label{bur-Tflat}
\end{eqnarray}

\section{Stabilization of sperical domain wall by charge}

The energy of an uncharged bubble forming from the BPS domain wall is
\begin{equation}
E_{0 bubble} = E_{wall} =
4\pi \int _0 ^\infty \rho r^2 dr \approx 4\pi r_0 ^2
\epsilon.
\label{E0tot}
\end{equation}
However, the Tolman mass
$M =\int dx^3 \sqrt{-g}(-T_0^0+T_1^1 +T_2^2 +T_3^3)$,
 taking into account tangential stress of the wall, is
negative
\begin{equation}
M_{Tolm. bubble} = - E_{wall} \approx -4\pi r_0 ^2 \epsilon.
\label{M0tot}
\end{equation}
It shows that the uncharged bubbles are unstable and
form  the time-dependent states \cite{CvSol}.
\par
Charged bubbles have extra
contribution caused by the energy and mass of the external electromagnetic
field
\begin{equation}
E_{e.m.} = M_{e.m.} = \frac{e^2}{2r_0},
\label{EMem}
\end{equation}
and contribution to mass caused by gravitational field of the external
electromagnetic field ( determined by Tolman relation for
the external e.m. field)
\begin{equation}
M_{grav. e.m.} =  E_{e.m.} = \frac{e^2}{2r_0}.
\label{Mgrem}
\end{equation}
As a result the total energy for charged bubble is
\begin{equation}
E_{tot.bubble} = E_{wall} + E_{e.m.} = 4\pi r_0 ^2 \epsilon
+ \frac{e^2}{2r_0},
\label{Etot}
\end{equation}
and the total mass will be
\begin{equation}
M_{tot.bubble} = M_{0 bubble} + M_{e.m.} + M_{grav.e.m.} =
- E_{wall} + 2E_{e.m.} = -4\pi r_0 ^2 \epsilon + \frac{e^2}{r_0}.
\label{Mtot}
\end{equation}
Minimum of the total energy is achieved by
\begin{equation}
r_0=(\frac{e^2}{16\pi \epsilon_{min}})^{1/3},
\label{rmin}
\end{equation}
which yields the following expressions for total mass and energy of
the stationary state
\begin{equation}
M_{tot}^* =E_{tot}^* =\frac {3e^2}{4r_0}.
\label{EMst}
\end{equation}
One sees that the resulting total mass of charged bubble is positive,
however, due to negative contribution of $M_{0 bubble}$ it can be lower
than BPS energy bound of the domain wall forming this bubble.
This is a remarkable property of the bubble models, existence the
`ultra-extreme' states \cite {CvSol} ) gives a hope to overcome
BPS bound and get the ratio $m^2 \ll e^2$ which is
necessary for particle-like models.

\section{Peculiarities of the rotating model and role of supergravity}

 For the rotating Kerr case  $J=ma$ and for $J\sim 1$ one finds out that
parameter $a\sim 1/m$ has Compton size.
Coordinate $r$ is an oblate spheroidal coordinate, and matter is foliated on
the rotating ellipsoidal layers \cite{Bag}. Curvature of space is
concentrated in equatorial plane, near the former singular ring, forming a
stringlike tube. \footnote{For the parameters of electron the phase
transition region represents an oblate rotating disk of Compton size
and thickness $\sim e^2/2m$.}

In supergravity, for strong fields there is also an extra contribution to
stress-energy tensor leading to negative cosmological constant
$\Lambda =-3 k^4 e^{k^2K}\vert W \vert ^2$ which can yield AdS space-time
for the bag interior.

The considered here supersymmetric model is more complicated than the
traditionally used domain wall models \cite {CvSol}, and it demonstrates
some new properties.
One of the peculiarities of this model is the presence of gauge fields
which, as it was shown in thin wall approximation, allow one to stabilize
 bubble to a finite size. Second peculiarity is the presence of a few
chiral superfields that can give a nontrivial sense to K\"ahler metric
$ K^{i\bar j}$ of the supergravity field models. One can expect that
extra degrees of freedom of the K\"ahler metric can play essential role
for formation of the bent (spherical or ellipsoidal) domain wall
configurations. In this case there appears a singularity in the K\"ahler
potential, and involving the axion and dilaton fields (coming from
low energy string theory) can be necessary to suppress its influence.
Therefore, some extra internal structure ("stringy or dilatonic core")
can appear for the bent domain walls on the Plankian scale. This second core
has to be placed inside of the Compton scale region connected
with the above domain wall structure of the chiral (Higgs fields).
It should be noted that the typical superstring BH solutions do not contain
the Higgs fields and are regular if only they are magnetically charged.
On the other hand, the close analogy of the superstring BH's and domain wall
solutions was mentioned in both approaches \cite{CvSol} (in particular they
have similar causal structure).
Both these types of models can be described by diverse versions of
supergravity. It leads us to assumption that the models of black holes and
domain walls are complimentary in the sense that the prospective exact
regular solutions has to be of a hybrid form containing an electrically charged
domain wall with a phase transition to core region described by a
magnetically charged BH solution to superstring theory.

We would also like to note, that the connected with superconductivity
chiral fields acquire a nontrivial geometrical interpretation in the
Seiberg--Witten theory and in the Landau -- Ginzburg
theory where the chiral superfields  refer to the moduli of the internal
Calabi--Yau spaces \cite{GVW}. In the case of a few chiral fields it gives
an interesting link to higher dimensions with an alternative look on the
problem of compactification.

\section{Conclusion}

The treatment shows that:
\begin{itemize}
\item
 in spite of the extreme smallness of the local gravitational
field supergravity can control the position of phase transition
at unexpectedly large distances;
\item
core of the Kerr spinning particle has the shape of oblate rotating disk
(of Compton size), and one can expect a sensitivity of differential sections
for polarized spinning particles depending on the direction of polarization.
\end{itemize}

It should be noted, however, that the parameters of elementary particles are
very far from the typical extreme BPS-states ($m \ll e$), and quantum
corrections for particles can be very high leading to a strong smearing of
the shape.

The model suggests that the formation of spin can be connected with
a nontrivial rotating vacuum state forming a disklike bag.
Since gravitational field is extremely small, its influence on the
geodesic motion of the scattering particles has to be negligible besides
very thin region (string) near the border of the Kerr disk where the strong
fields are concentrated. Vacuum state has almost lightlike boost in this
region.  The trapped partons are also relativistically boosted that
resembles Zitterbewegung and some old models of spin built
of the lightlike ring currents (H. H\"onl, A. Schild and others).

Thus, interaction by collisions
can occur only by direct contact with this thin string or with partons
inhabitting the bag either via the Coulomb excitation (including the case of
very soft photons).
Apparently, one can expect also excitation of the vacuum state of the bag
(Higgs fields) in the deep inelastic processes.

\bigskip

{\small We would like to thank Organizers of both the Conferences
 for kind invitation to give this talk and for financial support.}

\end{document}